# An interdisciplinary approach to certain fundamental issues in the fields of physics and biology: towards a Unified Theory

Fred H. Thaheld*
99 Cable Circle #20
Folsom, California 95630, USA

**Abstract**

Recent experiments appear to have revealed the possibility of the existence of quantum entanglement between spatially separated human subjects. And, in addition that a similar condition might exist between basins containing human neurons adhering to printed circuit boards. In both instances preliminary data indicates what appears to be nonlocal correlations between brain electrical activities in the case of the human subjects and also nonlocal correlations between neuronal basin electrical activities, implying entanglement at the macroscopic level. If the ongoing expanded research and the analysis of same continues to support this hypothesis, it may then make it possible to simultaneously address some of the fundamental problems facing us in both physics and biology through the adoption of an interdisciplinary empirical approach based on Bell's *experimental philosophy*, with the goal of unifying these two fields.

*E-mail address:* fthaheld@directcon.net

*We must think and act anew.*          Abraham Lincoln

## 1. Introduction

Interdisciplinary papers are always fraught with a certain element of danger in that it is very difficult to not only attempt to master two different subjects such as physics and biology but, to combine them in such a manner so that one can propose certain experiments in an effort to try to solve various longstanding fundamental issues confronting us in these fields not only individually but, collectively. This element of danger is exponentially increased when one is also hard pressed to satisfy the predispositions of the practitioners on both sides, many of whom are entrenched not only on their own turf but, look somewhat askance at the other's field when it begins to encroach in serious fashion on their own. So, it is a delicate balancing act, especially when one is attempting to simultaneously address what appear to be related but conflicting fundamental issues in the areas of quantum mechanics and nonlocality, relativity and locality, and biology.

Several authors have addressed the question of whether the concepts and methodology of physics and quantum mechanics could be applied in the field of biology and, a proposal has been made for a vehicle for founding physics on biology rather than the other way around (Matsuno, 1993), raising the possibility of the *biologicalization of physics*. In addition, the issue has been raised (Matsuno, Paton, 2000) of whether information is or could be processed at the quantum level in biological systems and, highlights two fundamental aspects of how to apply quantum mechanics to biology. One is the nature of time, whether local or global, synchronous or asynchronous and secondly, how to maintain a quantum coherence on a mesoscopic or even a macroscopic scale. The authors feel that 'biology is not about applying quantum mechanics as it is already known through the experience of traditional physics, but rather about an attempt to extend quantum mechanics in the manner that the physicists have not tried'. A similar theme is carried forward in varying detail in other papers dealing with the interconnection and interplay between quantum mechanical and biological processes (Matsuno, 1995; 2003).

Conrad (1989) has stressed, in his unified model of physics and biology, the possibility of a combination of two types of interactions involved in the formulation of quantum theory. The first being connected with force laws (such as the electromagnetic force) and reversible, and the second being associated with measurement and irreversible.

Josephson (2002) has explored this issue that quantum and biological accounts are complementary in the sense of Bohr (1958) and that quantum accounts may be incomplete. That 'biological and quantum accounts of nature might, like the wave and particle accounts of certain phenomena, be complementary rather than, as with the conventional view, entirely derivative of the latter'. The issue has also been raised (Josephson, Pallikari-Viras, 1991) that we may have to differentiate between two different types of nonlocality, one related to quantum mechanics and the other to what is termed *biological utilization of quantum nonlocality*. I have taken the liberty of reducing this terminology to *biological nonlocality* (Thaheld, 1999a; 2000b; 2003b). Additional related thoughts on this subject in a similar vein have also been advanced (Josephson, Rubik, 1992; Conrad et al, 1988).

Only a very few experiments have been conducted to date attempting to explore the possibility of a quantum physics-biology interrelationship, with the first one utilizing pairs of human subjects in Faraday cages, where just one of the pair is subjected to photostimulation, investigating possible electroencephalographic (EEG) correlations between human brains (Grinberg-Zylberbaum et al, 1994). Later experiments, building upon this pioneer research have been performed and continue to corroborate, with increasing experimental and statistical sophistication, these unusual EEG correlations (Standish, 2001; Richards et al, 2002; Standish et al, 2004; Wackermann et al, 2003).

Experiments have also been conducted which have revealed evidence of correlated functional magnetic resonance imaging (fMRI) signals between human brains (Standish et al, 2003). These correlations occurred while one subject was being photostimulated and the other subject was having an fMRI scan performed.

Research has also been ongoing for over a year at the Univ. of Milan (Pizzi et al, 2003; 2004a; 2004b; Thaheld, 2000a; 2004a) utilizing pairs of 2 cm dia. basins containing human neurons on printed circuit boards inside Faraday cages separated by 20 cm. Laser stimulation of just one of the basins reveals consistent wave-form autocorrelations between stimulated and nonstimulated basins. In addition, there are indications that biological quantum nonlocality has been observed in the coherence of induced magnetic dipoles involved in muscle contraction in single actin filaments at the mesoscopic level, (Matsuno, 2001; 2003; Hatori et al, 2002) and, that cell motility underlying muscle contractions is accompanied by a quantum mechanical coherence on a macroscopic scale (Matsuno, 1999; 2001). All these experiments are described in greater detail later in this paper and, when taken together, seem to be pointing us in an unusual direction implying entanglement and nonlocality.

I have already attempted to address, within an empirical setting in previous papers, this issue of applying the concepts and methodology of physics and quantum mechanics in the field of biology and vice versa. When I analyzed the first experiment in this field (Grinberg-Zylberbaum et al, 1994), and noticed that they had used plain photostimulation from a flashing strobe-light, I immediately suggested that checkerboard pattern-reversal photostimulation be used, which possesses several advantages over their technique (Thaheld, 1998; 1999a; 1999b; 2000b), in addition to several other types of stimulus proposals. All of the researchers mentioned previously that are involved in the human EEG research, have adopted this same experimental protocol with apparent success. In a further instance (Thaheld, 2000a) I proposed an experiment to determine if Einstein-Podolsky-Rosen (EPR) nonlocal correlations might exist between 2 neuron transistors and, as previously mentioned, a series of experiments have been performed for over a year at the Univ. of Milan (Pizzi et al, 2003; 2004b) which appear to be pointing in this direction regarding unusual electrical autocorrelations. In addition, another experiment has been proposed (Thaheld, 2001a) dealing with a possible resolution of the Schrödinger's cat paradox. This was recently explored, again at the Univ. of Milan, utilizing 2 spatially separated human neuronal basins (For greater detail see Sec. 5). Another experiment has been proposed (Thaheld, 2003a) to test Ghirardi's theory (Ghirardi, 1999) concerning possible photon superposition reduction mechanisms in perceptual processes which may be governed by nonlinear evolution laws. A modification of the original proposal is in the process of being formalized at the Univ. of Milan in anticipation of conducting a test of his theory.

As mentioned previously, I also introduced the concept of *biological nonlocality* (Thaheld, 2003b) and its possible relationship to the mind-brain interaction problem. And, both analyzed and proposed certain experiments in an attempt to resolve several outstanding issues concerning this problem. I would now like to expand upon this concept once again, utilizing what has become known as Bell's *experimental philosophy* (Whitaker, 1996), since the bulk of the experiments proposed in this paper rely very heavily upon the concept of nonlocality in general, and its specific adaptation to *biological nonlocality*. I find that this technique has a certain charm in that it allows one to both design, conduct and analyze experiments with a certain degree of clarity in this field which is normally missing in the present theoretical atmosphere.

Before we embark on this adventure, I would like to leave you with this caveat. The research group at the Univ. of Freiburg, dealing with the human subjects, stresses that "while no biophysical mechanism is presently known that could be responsible for the observed correlations between EEGs of two separated subjects, nothing in our results substantiates the hypothesis (Grinberg-Zylberbaum, 1994) of a direct quantum physical origin of correlations between EEGs of separated subjects" (Wackermann et al, 2003). And, the researchers at the Univ. of Milan state that, "Despite at this level of understanding, it is impossible to tell if the origin of this nonlocality is a genuine quantum effect, our experimental data seem to strongly suggest that biological systems present nonlocal properties not explainable by classical models" (Pizzi et al, 2003). I therefore take full responsibility for the conclusions I have drawn concerning *biological nonlocality* and, for the proposed experiments which follow.

2. **The Aspect EPR nonlocality experiment and its analogy to the Grinberg-Zylberbaum experiment**

The famous article by Einstein-Podolsky- Rosen (EPR), described a situation where two entangled quantum systems are measured at a distance (1935). Bell (1964) later showed that the correlation between the data cannot be explained by local hidden variables and, a first limited test of Bell's theorem with entangled photon pairs was successfully performed (Freedman and Clauser, 1972) with fixed polarizers (Fig. 1- This is a simplified schematic of the actual setup). Later, a more sophisticated experiment was conducted (Aspect et al, 1982), also with entangled photon pairs and using rapidly switchable random polarizers, which revealed once again and, with greater certainty, the existence of nonlocal polarization correlations between these entangled photons, while ruling out the possibility of any local hidden variables.

In this experiment, a polarization measurement on entangled photon v1 with polarizer I, as recorded by detector PM 1, resulted in a simultaneous transference of a similar polarization state or polarization information to photon v2, in nonlocal fashion as shown hypothetically by the wavy line. This transference effect is always and only derived when a polarization measurement is made on the photon v1 by polarizer I, i.e. when someone *consciously chooses* to place polarizers in photon v1's and v2's paths. The question of just how this information transfer is accomplished nonlocally between these photons and, just what information is transferred, remains one of the fundamental mysteries in quantum mechanics.

Just how might this quantum nonlocal experiment relate to present research involving what appears to be biologically entangled human subjects and human neuronal basins? A decade ago a research group led by Grinberg-Zylberbaum (1994) stimulated subjects (for convenience let us call them Alice) in a Faraday cage with photostimulation (Fig. 2) and elicited a Visual Evoked Potential (VEP) on their EEGs. Faraday cages are used to screen out nearly all relevant electromagnetic, acoustic and other influences but, not gravity. They noted that 25% of the time, what they called a Transferred Potential (TP) or simultaneous brain-wave events (which others prefer to call correlations or autocorrelations of electrical activities) appeared in the brain electrical activity or EEG of a nonstimulated subject (let us call him Bob) also in a Faraday cage, in what was referred to as EPR nonlocal fashion, as again shown by the hypothetical wavy line. The TP was not an exact duplicate of the VEP nor was any conscious subjective experience transferred between Alice and Bob, such as of the white flashing strobe light. Without photostimulation they noted that there was never any VEP or measurement on Alice's EEG and also, no TP or transfer of information on nonstimulated subject Bob's EEG, just like if one doesn't make any polarization measurement on the first photon v1 with polarizer I, no polarization information is ever sent to the second photon v2.

The source of this TP is the initial photostimulation of the first subject, which constitutes a measurement analogous to the photon polarization measurement, with a subsequent transfer of information from a stimulated subject Alice to a nonstimulated subject Bob. This same effect has also been seen by researchers at the Univ. of Washington (Standish, 2001; Richards et al, 2002; Standish et al, 2004) and, in more rigorous fashion, at the Univ. of Freiburg (Wackermann et al, 2003). They have also not seen any transfer of conscious subjective experience of the checkerboard pattern used to stimulate the subjects. Up to now the question of what might be the source of this TP and what the TP represents, has also remained a mystery. So, it appears that we now have two mysteries, like Chalmers has stated (1996), which appear to be one and the same and, now capable of being more readily subjected to experiment in the biological or biophysics realm (Thaheld, 2004a).

Even though the means of information transfer and the information that is transferred between the entangled photons and the entangled human subjects appears to spring from the same source and, is a direct result of some type of measurement, there are several major differences between the two entities which are very important and requires a shift in one's thinking. First, one is nonliving and usually involves just a pair of entangled photons or particles (excluding superconductors at this point), while the other is living and involves a large number of entangled neurons, both separately and collectively consisting of a massive and varied number of different types of molecules. Second, we know both states in advance whereas with entangled photons both states are completely unknown. Third, the data which is derived is not random as it is with the photons.

After a polarization measurement is made on the entangled photons, the wave-function collapses, entanglement is lost and, a new pair of photons has to be prepared for another measurement. In contrast, it appears that after one makes a measurement on entangled human subjects, that the macroscopic entangled living system, consisting of an immense number of diverse entangled entities, is able to constantly maintain or regenerate this entangled state for a long period of time and, in a very rapid fashion, as has been revealed in the Freiburg experiments (Wackermann et al, 2003). It would have been impossible for them to repeatedly photostimulate their subjects at a given Hz rate over a period of minutes, and also observe correlations of brain electrical activities with the nonstimulated subjects during this period of time, if entanglement was not being constantly maintained between both subjects. With the entangled living systems one is able to make repeated measurements with the *same systems*, which you cannot do with nonliving entangled photon, electron and other particle systems.

In addition, it appears that research now being conducted at the Univ. of Milan with human neural networks adhering to printed circuit boards (Fig. 3) (Thaheld, 2000a; 2004a; Pizzi et al, 2003; 2004b), may have also revealed this same phenomenon, only in a more striking fashion, due to the absence of the usual high EEG noise level associated with human subjects. Also, in the case of the entangled human subjects, one is comparing and analyzing simultaneous events on their EEGs in a very narrow range of from 2-8 µV, which makes the statistical analysis that much more difficult, requiring time-consuming averaging techniques.

With the entangled human neural networks, the wave-function amplitudes are in a range varying from 5-30 mV, so that one gets very superior and consistent electrical autocorrelations between the two entangled basins reducing the need for averaging techniques, due to the lack of the usual EEG noise. Again, and most importantly, it appears that entanglement is being maintained or regenerated at a very high Hz rate of stimulation and for long periods of time.

As I have pointed out recently (Thaheld, 2003b; 2004a), it may well be that what has been referred to as an immaterial or non-material mental event (as reflected in the wavy line of Fig. 2), which can interact with a material brain, resulting in a depolarizing event, is equivalent to the nonlocal quantum event (the wavy line in Fig. 1) which interacts with a photon causing a similar polarization event. My analysis of this whole situation has then led me to ask, what exactly do we mean when we use the term 'immaterial' with regards to the mind-brain interaction problem? I am bothered by the fact that when a polarization state is somehow or other transferred from one photon to another, physicists do not refer to this process as 'immaterial' (excluding Einstein's 'spooky action at a distance' comment!; Mermin,1998) but, as quantum mechanical in nature involving nonlocality. Why then in the field of biology, do we apply the term 'immaterial' to a mental state which seems to affect in similar fashion a material or brain state? Would it not be much simpler to just drop this 'immaterial' term and refer to this process in the future as quantum mechanical nonlocality or *biological nonlocality?* Are we not losing sight some of the time of the main issues confronting us due to the very nature of semantics?

Another major difference is that with the living systems we may now be able to examine exactly where and how the transition is made from a mental or quantum state to a brain or classical state and vice versa. This is like asking the question, '*What and where is the biological equivalent in a neuron of the polarizer used in physics and can this be considered as a 2-way transducer?* Several possible locations have been postulated in the neurons which could be the targets for immaterial mental events. In this regard Eccles (1994) has postulated that a non-material mental event could influence the subtle probabilistic operations of synaptic boutons, focusing attention on the effective structure of each bouton, which are the paracrystalline presynaptic vesicular grids, as the targets for non-material mental events.

One will now have many degrees of experimental freedom available in biological systems such as neurons to explore these possibilities, which are denied one when you are limited to just nonliving photons, electrons, etc., where collapse is complete and irreversible each time and, usually only involving a pair of entangled particles at that (Thaheld, 2003b).

One of the reviewers of this paper pointed out that quantum mechanics has its own nonlocality as demonstrated in an EPR-type experiment however, when mental events are claimed to be nonlocal, what would that mean? And, if mental events are of quantum mechanical origin, how could it be demonstrated experimentally? I would like to respond to these questions in the following manner.

First, when the term 'mental event' (or conscious subjective experience) is used in this paper, it refers not only to events taking place within and between the brains of two human subjects but also, between two basins containing human neurons adhering to printed circuit boards. I feel that the initial experiment by Grinberg-Zylberbaum (1994) and the corroborating experiments at the Univ. of Washington (Standish et al, 2004) and the Univ. of Freiburg (Wackermann et al, 2003) with spatially separated human subjects, both of whom are in Faraday cages, should essentially screen out any neural, electromagnetic, acoustic or electrolytic influences, leaving only mental events as one of the main processes possibly involved. In addition, ongoing experiments at the Univ. of Milan (Pizzi et al 2003; 2004a; 2004b) with 2 separated human neuronal basins, also in Faraday cages, are analogous to the Freedman-Clauser nonlocality experiment (1978) where the polarizers were held in a fixed position. Since one knows that Freedman-Clauser provides direct evidence for a quantum nonlocal effect between two spatially separated and entangled photons, then one should be justified in saying that this same quantum nonlocal effect may be utilized also between, not only entangled human subjects but, entangled human neuronal basins. And, that this can be referred to as a mental event (since neurons are involved in both instances) which is equivalent to the information transferred in correlated photon polarization events.

As per Libet (1994) we know that neural events can influence, control and presumably 'produce' mental events, including conscious ones. The reverse of this, that mental processes can influence or control neuronal ones, is known as the 'reverse direction' problem. In his experimentally testable solution he has proposed a conscious mental field (CMF), whose chief attribute would be that of a unified conscious subjective experience, with the second attribute being a causal ability to affect or alter neuronal function. He states that this CMF would not be in any category of known physical fields, is not describable in terms of any known physical theory and, is only detectable in terms of the subjective experience reported on by an individual subject. One is immediately restricted by having to define some new 'field', which supposedly does not now exist and, in having to rely upon anecdotal material. He makes no mention of a quantum mechanical nonlocality possibility.

In all the experiments performed to date (Sec. 2), there was no transference of conscious subjective experience reported. However, all these experiments revealed that photostimulation of the brains of human subjects, which resulted in specific neural events or visual evoked potentials (VEPs), simultaneously caused neural events to occur in nonstimulated subjects. The same effect happens also between human neurons adhering to printed circuit boards. Now, since one cannot say that neural events were transferred between these entities (as they were all inside Faraday cages), since this would violate both special relativity and the conservation of energy laws, you are left with only mental events or the 'reverse direction'. In addition, if any of the several techniques I have proposed to attempt to rectify this matter of transferring and detecting conscious subjective experience between human subjects is successful (Thaheld, 2003b), one may be able to resolve both issues simultaneously via the quantum mechanical nonlocality route.

We have no trouble in saying in physics that the same quantum nonlocal effect is involved in both entangled photons' correlated polarization measurements as is present between correlated electrons' spin measurements, even though the photon's nonlocal polarization correlation is brought about by polarizers and the electron's nonlocal spin correlations are brought about by inhomogenous magnetic fields in a Stern-Gerlach device. Furthermore, that this quantum nonlocal effect acts upon *both* particles equally, even though the photon possesses no mass while the electron does possess mass. We do not attempt to conjure up another field or force for this mutual effect involving both photons and electrons, so why then would one feel compelled to think that this same quantum nonlocal effect could not also be utilized in biology, as it is so successfully used and delineated repeatedly in physics? I also touch upon this matter further in Sec. 4.

Before leaving this subject I would like to briefly comment on why I used the names of Alice and Bob in this section. Alice and Bob are the archetypal individuals who have been used in a number of papers discussing teleportation of an unknown quantum state (Bennett et al, 1993), quantum communication (Bennett and Wiesner, 1992) and quantum cryptography (Ekert, 1991). This requires the production of polarization-entangled states of photons (Kwiat et al, 1995) which are then manipulated in various ways by Alice and Bob depending upon the final end result to be achieved. In addition to sharing a series of EPR-correlated pairs of photons and a channel capable of carrying classical messages, they are required to make joint measurements of their photons and send the classical results of these measurements to each other.

It then occurred to me that, based upon the recent experiments with what appear to be entangled human subjects (Wackermann et al, 2003; Standish et al, 2004), that since Alice and Bob can also be considered to be entangled, they might no longer need to generate nor manipulate entangled EPR photon beams and, that the transmission of information between them can now be moved to an entirely new level. They are now in a position to not only manipulate themselves via photostimulation and their EEGs but, each other in 2-way fashion. Furthermore, even if they are not entangled themselves they can use entangled human neuronal basins as their proxies, labeling them as Alice and Bob. So it is that I elected to elevate Alice and Bob into the positions formerly occupied by the entangled photons which were once subject to their manipulation! What I find most fascinating is that Alice and Bob have in effect now become *both the measuring devices* and the *measured systems*, or *the observer* and *the observed*, as have the entangled neuronal basins! The noted neurosurgeon Wilder Penfield addressed the problem associated with this same concept decades ago when he pondered the prospect of performing brain surgery on himself by asking, "Where is the subject and where is the object if you are operating on your own brain?" (Penfield, 1976; Goswami, 1993).

And finally, this raises a very interesting question as to whether, if Alice and Bob are identical twins and entangled, Einstein's 'twin paradox' (Uzan et al, 2000), where Bob makes a trip in a noninertial frame with a time-varying velocity and returns to earth younger than Alice, who has remained on earth in an inertial frame, would still be correct?  Especially since this entanglement involves nonlocality between subjects in two different frames.

3.  **Are there two different types of decoherence or wave-function collapse, one for living and one for nonliving systems?**

One then has to logically ask the question, based upon the previous analysis as to whether, in the case of the living system, collapse is complete each time, as is the case with the entangled photons or, it is only partial and entanglement is maintained or constantly regenerated, whichever case applies?  That living systems are able to rapidly oscillate from linear to nonlinear and back to linear in rapid, regenerative fashion?  If one looks at a typical neuronal action potential (Fig. 4), the membrane is depolarized from a resting level of -40 mV to a level of +40 mV, so that there is a 'reversal' of the membrane potential.  And, in less than 1 ms the membrane returns to its previous resting level of -40 mV, waiting for the next depolarization event or, if you will, the next measurement.

This makes me think that this is why entanglement is either never lost or is constantly maintained or regenerated in a living system and, that we may be looking at two different types of collapse.  One complete and irreversible in the case of photons, where just a few entities are entangled and, one complete or partial but, repetitive and reversible in the case of a neuronal system, consisting of an immense number of entangled living entities. In addition, this also indicates that the neuronal system is able to resist decoherence and that there are two different types of decoherence processes, one for living and one for nonliving entities.

This line of reasoning appears to be further borne out by the recent research of Matsuno (2001; 2003; Hatori and Matsuno, 2002), indicating that an extremely minute aspect of biological quantum nonlocality has been observed in the coherence of induced magnetic dipoles in muscle contraction involving single actin filaments and proposes (Matsuno, 1999) a condensation of the atomic degrees of freedom constituting the filament into a macroscopic quantum state carrying a nonvanishing linear momentum. This means that a very unusual new level of experimental flexibility can now be introduced into interdisciplinary research in the area of biophysics, which is denied one if you limit yourself to only physics or biology as separate fields.

**4. Anesthetics, neurons and the matter of coherence or entanglement**

It will now be possible to subject both the macroscopically entangled human subjects (Fig. 2) and the human neuronal basins (Fig. 3) to actual experiments utilizing anesthetics to see what effect they might have on entanglement, nonlocality and quantum coherence (Thaheld, 2004a).

1. A general anesthetic is administered to Alice inside a Faraday cage (Fig. 2) and, after the loss of consciousness, she is subjected to a flash stimulus and a VEP is recorded on her EEG. It is very important to note here that clinical electrophysiological brain monitoring shows reduction and desynchronization of the brain neural level dynamics during general anesthesia (Hameroff, 1998). Now, under conditions like this, will a TP or electrical correlation still be observed in the EEG of the 2$^{nd}$ nonstimulated subject Bob just like before? If it isn't, this will show that even though there was a VEP, the anesthetic did something to the quantum coherence or nonlocal transference of information between the two subjects. I.e., entanglement was broken somehow by the anesthetic, which might tie in indirectly with Hameroff's theory (1994) that microtubule based quantum coherence should be sensitive to general anesthesia. If however, a TP is elicited in the EEG of the nonstimulated subject Bob, we then administer a general anesthetic to Bob and repeat the flash stimulation of Alice, after allowing her general anesthetic to dissipate, and see if a TP is still received by Bob. The results of this experiment will tell us something very fundamental about the role anesthetics play with regards to the maintenance of quantum coherence or entanglement between these human subjects and, if entanglement is reestablished following dissipation of the anesthetic.

2. To further refine this process, an anesthetic can be administered to a Univ. of Milan neuronal basin A (Fig. 3) and, it is then subjected to the usual laser stimulation (Thaheld, 2004a). First, we want to see if a series of wave-forms is still generated as it was before use of this anesthetic (allowing for reduction and/or desynchronization of the wave-forms or action potentials) and, if it is, we then want to determine if autocorrelation of the electrical activity is still observed between stimulated basin A and nonstimulated basin B in nonlocal fashion as was previously observed. If not, this will tell us that entanglement was broken or disrupted by the anesthetic. We then want to see if, after allowing the anesthetic to dissipate in neuronal basin A, when we once again laser stimulate A, does the resulting wave-form become autocorrelated with that of nonstimulated neuronal basin B once again i.e., is entanglement reestablished even though the 2 neuronal basins are not in direct contact! Either entanglement was never lost or, it was lost and then reestablished on a very rapid basis, even though the two neuronal basins were 20 cm apart. One can also reverse this situation and anesthetize basin B and see if entanglement and autocorrelation is lost. The results of this experiment would have a more direct bearing on Hameroff's microtubule theory concerning quantum coherence (1994) and, his quantum hypothesis of anesthetic action (Hameroff, 1998), implying a very robust coherence with quantum nonlocal interconnections. Furthermore, by being robust it allows the microtubules to resist normal environmental decoherence for long periods of time and one can then prove the existence of this robustness via constant laser stimulation! If it were not robust this effect would disappear extremely rapidly when subjected to a constant stimulus of this nature.

3. Of course, I would be remiss if I didn't mention one of nature's own ultimate and random anesthetic, what we refer to as *temperature*. As we know from experience, a very few degrees higher or lower for the human subject and consciousness is either lost or severely impaired. This may enable us to vary the temperature in either direction for either humans or the neuronal basins and determine if entanglement or quantum coherence is lost.

In a recent paper (Thaheld, 2003b), I made an argument for what I have already referred to as *biological nonlocality*, based on the fact that both the human subjects (Fig. 2) and the human neuronal networks on printed circuit boards (Fig. 3) are all inside Faraday cages, thereby enabling us to selectively rule out all neural, acoustic, electrolytic and most electromagnetic means of communication or correlation but, not gravity, nor by inference, quantum gravity (Thaheld, 2004a). This gravitational aspect seems to be very important since both photons without mass and electrons and other particles with mass are equally affected by gravity, just as they are equally affected by entanglement and nonlocality. The major weakness in this analysis is that due to the close proximity of the human subjects as measured in m, and that of the neuronal basins as measured in cm, local hidden variable effects cannot be ruled out due to the speed of light.

5. **A proposed resolution of the Schrödinger's cat and Wigner's friend paradoxes**

The possible states of a system can be characterized by state vectors, also known as wave-functions which change in two ways, continuously as a result of a passage of time and discontinuously if a measurement is carried out on a system (von Neumann, 1955; Wigner, 1963; Shimony, 1963). This second type of discontinuous change, called the reduction of the state vector or collapse of the wave-function, is unacceptable to many physicists. Based upon his analysis of the EPR paper (1935), Schrödinger found a way to make the measurement problem too obvious to be missed by introducing what has become well known as Schrödinger's cat (1935). Let us quote him directly upon this subject.

"One can set up quite ridiculous cases. A cat is penned up in a steel chamber, along with the following diabolical device (which must be secured against direct interference by the cat): in a Geiger counter there is a tiny bit of radioactive substance, so small, that perhaps in the course of an hour one of the atoms decays but also, with equal probability, perhaps none; if it happens, the counter tube discharges and through a relay releases a hammer which shatters a small flask of hydrocyanic acid. If one has left this entire system to itself for an hour, one would say that the cat still lives if meanwhile no atom has decayed. The first atomic decay would have poisoned it. The wave-function of the entire system would express this by having in it the living and the dead cat (pardon the expression) mixed or smeared out in equal parts". I.e., if the half-life of the nucleus is one hour, then after this time the correct quantum mechanical description of its state is that it is neither definitely decayed nor definitely non-decayed, but rather in a linear superposition of two states. Wigner later expanded upon this concept by introducing what is known as Wigner's friend (Wigner, 1961).

At first blush you may be wondering why I would even want to discuss Schrödinger's cat, as Schrödinger himself introduced his cat proposal with the words, "One can even set up quite ridiculous cases", while other physicists refer to it as bizarre or a joke, since some feel that entanglement is highly vulnerable to environmental decoherence processes (Tegmark, 2000), while others appear to have successfully rebuted his arguments (Hagan et al, 2002). All this changes when we move from the realm of physics and quantum mechanics to biophysics. In the case of certain living entities it now appears that entanglement can be maintained between them *as long as they remain alive*, which moves the cat matter beyond the ridiculous or bizarre to a firmer footing.

Several years ago I proposed a revised experiment (Thaheld, 2001a) in an attempt to resolve the Schrödinger's cat and Wigner's friend paradoxes (Fig. 5) without ever once opening up the box, in the following fashion. A radioactive source is placed in a Faraday cage with a cat (if need be for company) and Wigner's friend Alice, who will be entangled with an outside observer Bob. Sometime during the course of an hour it will decay randomly and will trigger photostimulation of Alice at a given Hz rate, with the result that a VEP is generated in her EEG, while a simultaneous TP is elicited in the EEG of the nonstimulated subject Bob, also in a Faraday cage. Alice and Bob will have agreed in advance that when he observes a certain sequence of events on his EEG, that this will be an indication to him that the atom has decayed. This means that *until* he sees this TP on his EEG, he has known second by second that there has been no decay of the atom and, the life or death status of the subjects involved, without ever opening up the cage by just observing his EEG record. And, if I am correct in my assumption, that we are looking at a new concept of *nonlocal knowledge*, or knowledge conveyed in nonlocal vs classical fashion, which will enable one to always know what is going on or has taken place inside a closed box, without ever making any measurements or observations in a *classical* manner, i.e. by opening the box.

If the radioactive decay does not take place, as revealed by *no specific activity* showing up on the outside observer's EEG, this will constitute a nonlocal measurement instant by instant by the outside entangled observer. Seeing *nothing* on his EEG has told him *something*. And, if and when the radioactive decay does take place, as revealed by specific activity showing up on his EEG, this will once again constitute a nonlocal measurement. Seeing *something* on his EEG has told him *something*.

The question has to be raised as to whether the presence or absence of *nonlocal knowledge* would constitute a measurement and observation by itself? This concept is applicable to not only Schrödinger's cat or Wigner's friend type situations but, any other activities which might take place in a closed box, where inside and outside observers are entangled and the inside observer is also in a state of superposition with a random stimulus device. In addition, if recently proposed animal nonlocality experiments are successful (Thaheld, 2004b), we may even be able someday to use entangled cats where just one of them is photostimulated.

I am sure that at this juncture some of you are already saying "Hey, wait a minute. Fred has made a mistake and is mixing apples and oranges, superposed and entangled states at the same time, whereas Schrődinger stressed just the superposed aspect". This was pointed out to me several times after I devised this approach and, I must plead guilty to the charge. I have violated Schrődinger's fundamental dictum regarding "leaving this entire system to itself for an hour", by mixing in items with the cat or Wigner's friend, which are entangled with the *outside*, such as with other cats or humans. However, in light of the recent experiments revealing the possibility of entanglement between human subjects and human neuronal basins and, proposed animal entanglement experiments (Thaheld, 2004b), the possibility arises that there may be occasions when one *cannot* "leave this entire system to itself" (in a state of superposition), because it is naturally entangled and, that Schrődinger's original proposal may therefore not be completely correct.

A proposal has also been made to perform similar Schrödinger cat-type experiments by using two Univ. of Milan entangled human neuronal basins (Thaheld, 2000a, 2004a; Pizzi et al, 2003), and just such a series of experiments was recently performed at the University of Milan. A radioactive source triggered a laser in random fashion in neuronal basin A (Fig. 3), leading to stimulation of some of these neurons. First, by merely observing the lack of any electrical autocorrelation signals from the nonstimulated neuronal basin B only, they are able to tell instant by instant exactly what is going on inside the Faraday cage containing neuronal basin A, without ever opening up this cage, based upon a *non nonlocal* response. And, they know the exact moment when the random event takes place since the resulting laser stimulation caused basin A to fire, resulting in a simultaneous autocorrelated nonlocal response from basin B.

We have only very recently and, for the first time, been able to realize the thought experiment of Schrödinger concerning what is referred to as a quantum superposition of truly macroscopically distinct states, utilizing superconductors and superconducting quantum interference devices or SQUIDs (Friedman et al, 2000; van der Wal et al, 2000). This was based upon proposals that under certain conditions, a macroscopic object with many microscopic degrees of freedom could behave quantum mechanically provided that it was sufficiently decoupled from its environment (Caldeira and Leggett, 1981; Leggett et al, 1987; Weiss et al, 1987). In order to accomplish this decoupling, it is necessary to drastically reduce the temperature down to around 8 K. It appears that we will now be able to move this whole process, in what must look like a miracle, to the observable macroscopic biological level, in seeming contradiction of quantum mechanics and relativity, since entanglement will not only be taking place at around 300 K but, it appears that the living system will be able to remain decoupled from the environment for a long period of time, *as long as the neurons remain alive.* Thus it is that a *living* macroscopic system with many microscopic degrees of freedom (many orders of magnitude more than are observed in the case of superconductors) may be able to behave in a quantum mechanical coherent fashion since it is sufficiently and continuously decoupled from the environment. Or, that biological coherence or entanglement may feed on decoherence! This aspect is quite different compared to physical coherence or entanglement competing with decoherence (Matsuno, 2004).

# 6. Superluminal communication?: The matter of uncontrollable vs controllable nonlocality

Relativity theory postulates the nonexistence of faster-than-light 'signals' but, does not necessarily impose an analogous requirement upon all other conceivable kinds of influences (Stapp, 1988). He proposes the existence of superluminal 'influences' between the entangled photons which are not considered to be 'signals', with the result that no conflict with the theory of relativity is entailed.

Based upon what is known as the Eberhard theorem (1978) it is felt that no information can be transferred via quantum nonlocality. Shimony (1984) has also come to the conclusion that the nonlocality of quantum mechanics is uncontrollable and, cannot be used for the purpose of sending a signal faster than light. In this sense he feels that there may be a 'peaceful coexistence' between quantum mechanics and relativity theory.

Why is it that the entangled photons in the Freedman-Clauser (1972) experiment cannot be used to transmit information faster than light? After all, it has been shown that nonlocal correlations exist between these photons and one would logically think that you should be able to perform this feat. The polarization correlations observed in this experiment cannot be used to transmit information faster than light because they can only be detected when the statistics from the measurements on each side are compared in a classical fashion which is dependent upon the low efficiency of the detectors. The act of polarization measurement on photon v1 forces it to move from the quantum indeterminate level to a specific, determinate level and it is this information that is transmitted nonlocally to entangled photon v2. When Alice, let us say, measures the polarization of photon v1 along a direction she chooses (horizontal or vertical) she *cannot choose the result*, nor can she predict what it will be.

Once she makes her measurement, Bob's photon v2 simultaneously receives nonlocal information regarding a similar state of polarization where he *cannot choose the result either*. Since Alice has no control over the results she gets, she can't send any meaningful information of her own to Bob. Similarly, Bob can choose one of several polarization measurements to make but, he will not know the result ahead of time and cannot receive any information. Alice and Bob can only see the coincidence of their results after comparing them using a conventional method of communication which does not send information faster than light.

I feel that while this reasoning is correct for nonliving systems such as photons where, once a measurement has been made, the wave-function collapses and they become disentangled, the situation is much different for *entangled living systems* such as the human subjects and the human neurons on printed circuit boards. Here, instead of being limited to just a pair of entangled photons at the microscopic level, which only very briefly spring into existence and then are gone, we are looking at a massive number of living entangled particles at the macroscopic level, either resisting the usual decoherence faced by the photons or maintaining or regenerating entanglement, thereby enabling us to achieve controllable superluminal communication (Thaheld, 2000a; 2001b; 2003b).

Why is it that one might be able to achieve superluminal communication by utilizing living entangled entities? You will recall that in the original Grinberg-Zylberbaum experiments (1994) photostimulation was utilized to elicit a VEP in the brain waves of the stimulated subject. In this paper it was mentioned that if a flickering light signal were used the normal VEP often carries a frequency signature. Goswami has proposed (Grinberg-Zylberbaum, 1994) that to the extent that this frequency signature is also retained in the TP, it may be possible to send a message, at least in principle, using a Morse code.  I.e., one could vary the frequency of photostimulation to resemble a code.  He has further suggested that the brain obeys a nonlinear Schrödinger equation in order to include self-reference (Mitchell and Goswami, 1992) and, that for such systems, message transfer via EPR correlation is permissible (Polchinski, 1991).

  The problem is that although the elicited VEP in the stimulated subject's EEG carries a distinct amplitude and frequency, it is in a very low µV range and, the µV of the TP in the nonstimulated subject's EEG is not only several orders of magnitude lower but, his wave-form as regards both frequency and amplitude, comes no where near to being a copy of the original VEP.  That the frequency signature is not retained in the TP represents a serious defect analogous to the low detector efficiencies.  Thus, one can perceive some type of very rudimentary message or code, only after averaging over many cycles of photostimulation, which is a lengthy process.  Recently experimenters at the Univ. of Freiburg (Wackermann et al, 2003) achieved a more robust replication of the Grinberg-Zylberbaum research using patterned photostimulation.  However, they also noted this same problem, although in more dramatic fashion, with an averaged VEP peak amplitude of around 8µV and the averaged correlated peak amplitude of less than 2 µV, with no wave-form similarity between the initiating VEP and the resulting correlated electrical activity.  Their results indicate a high co-incidence of variations or correlations of the brain electrical activity in the nonstimulated subjects with brain electrical responses of the stimulated subjects.  They did not see any VEP-like wave-forms in the averaged EEG of the nonstimulated subjects.

  The major difference between the living and nonliving systems is that when we make a measurement, in contrast to Alice with her photons, we are able to *choose the result* for both Alice and Bob, who remain entangled after many repeated measurements and, we know in advance what results we will get! In addition we will be able to *choose these results* in a 2-way communication fashion by stimulating (measuring) either one of the parties at our option. Furthermore, Alice and Bob, since they are entangled, will now be able to *choose* their own types of measurements and results. They must simply agree in advance what the signals they will be seeing on each others' EEGs translate into, thereby achieving superluminal communication without any need to compare both of their results after the fact, since they will now be making their comparisons in nonlocal fashion.

At the present time it appears that it is going to be much simpler for Alice and Bob to use the entangled human neuronal basins for superluminal communication, since the wave-form signature generated by basin A is more fully correlated with that of nonstimulated basin B (Fig. 3), although at a lower amplitude. Both the laser stimulation rate of basin A and the nonstimulated electrical correlated rate for basin B approach very close to 100% each time due to the large number of photons in the laser beam used to activate some of the neurons, which can vary in number from 10-30 sometimes, up to 100-300 at other times (Pizzi, 2004b). So, as detectors, these neurons are almost 100% efficient.

You know that you have achieved superluminal communication, even though the 2 basins are only separated by 20 cm, if you are able to exchange information in a nonclassical manner in a 2-way fashion. I.e., the two observers Alice and Bob, stimulating and monitoring their own respective entangled basins, never have to get together to compare their separate results after the fact and in a classical manner, as is the case with the correlated photon experiments. This is because these two observers with their two entangled basins can communicate their results of what they received from each other back and forth in a 2-way nonlocal fashion. The statistics from the measurements on each side can be compared in a nonlocal fashion! And, they know that no conceivable classical means of communication or information exchange has been relied upon to achieve this communication, since one by one most of the conceivable locality loopholes can be identified and closed, leaving one with nonlocality as the most probable answer.

The unusual thing about these neuronal basins is that they also generate spontaneous action potentials without any laser stimulus. At the present time the researchers at the Univ. of Milan are unable to record action potentials from single neurons (Pizzi et al, 2004b). In the future this may reveal that the basins are able to stimulate each other in nonlocal fashion, making their own measurements or *autodecoherence* without any outside source such as the laser being required and, in 2-way communication! Since they are supposedly not conscious this would raise some interesting philosophical questions. It is of interest to note here that the researchers from the Univ. of Milan presented a paper (Pizzi et al, 2004b) at a recent SPIE symposium on defense and security that has implications for superluminal communication in the areas of submarines, spacecraft, etc. While I realize that this proposal to achieve superluminal communication appears to conflict with special relativity and causality (Stapp, 1988; 1994; 1997), this is a whole other issue way beyond the scope of this paper.

7. **A new technique for resolving the Bell inequality loopholes utilizing human neuronal basins**

As has already been pointed out in Sec. 2, Einstein-Podolsky-Rosen described a situation where two entangled quantum systems are measured at a distance. Bell hypothesized (1964) that the correlation between the data cannot be explained by local variables. This has become known as Bell's inequality. All the experiments performed to date to test Bell's theory (Freedman et al, 1972; Aspect et al, 1982; Kwiat et al, 1995) confirm the predictions of quantum mechanics with regards to entanglement and nonlocality. However, the above experiments don't succeed in completely ruling out a local realistic explanation because of two essential loopholes (Vaidman, 2001).

The first loophole has to do with detector efficiency, in that only a small fraction of all photon pairs which are generated are detected (Pearle, 1970). One then has to assume that the pairs registered are a fair sample of all pairs emitted (Weihs et al, 1998). This loophole has been closed experimentally by using two massive entangled trapped ions (Rowe et al, 2001) but, it does not simultaneously close the locality loophole.

The second loophole, that of locality, states that whenever measurements are performed on two spatially separated particles, any possibility of signals propagating with a speed equal or less than the velocity of light between the two parts of the apparatus must be excluded (He et al, 2003). This loophole was closed in two recent experiments (Weihs et al, 1998; Tittel et al, 1998). It is now obvious that to close both loopholes simultaneously in the same experiment still remains a challenge (He et al, 2003). I believe that there may be a way simultaneously around both of these problems if we utilize two entangled human neuronal basins in the following manner.

First, it is necessary to visualize that the entangled photon pairs (Fig. 1) which fly apart at the speed of light, are to be replaced with two static macroscopically entangled human neuronal basins (Fig. 3) separated by 400 m, so that light takes approximately 1.3 µs to travel between them. This means that there will be no need for polarizers and no need to switch them while the photons are in flight, making the experimental protocol that much simpler. At the present time the neuronal basins at the Univ. of Milan are only separated by 20 cm, so it is unknown if entanglement will persist out to 400 m or further although, if this is truly a nonlocal phenomenon, distance should have no effect. At this distance of 20 cm it appears that the wave-forms between the stimulated and nonstimulated basins are simultaneous and autocorrelated, since a reaction lag in ms should be visible, which they say it is not. The researchers feel that if a lag exists, it is rather in the range of µs. If the experiment reveals that the lag is less than 1.3 µs between the two neuronal basins, this would be evidence for nonlocality and represent a closure of the locality loophole. If it is greater than this we would have to separate the basins by a greater distance.

Second, this means that not only have the two neuronal basins been substituted for and become in effect analogous to the entangled photons but, they have also become their own entangled detectors! These neuronal basins have a detection efficiency approaching almost 100% by virtue of some of the neurons in the 2 cm stimulated basin being bombarded by the 630 nm laser pulse with a 4 mm dia over a period of either 1 ms or 300 ms per cycle, causing a large number of action potentials to sum each time and resulting in a series of highly specific wave-forms. For every electrical wave-form generated by the laser, you get a simultaneous elicited autocorrelated wave-form in the nonstimulated basin, with both of them possessing different amplitudes varying from 5-30 mV and, without the usual EEG clutter or noise. This can then become a 2-way Bell test by alternately stimulating either basin which one cannot do with photons! Adoption of this technique might provide us with an alternate approach to simultaneously resolving these 2 loopholes.

## 8. Conclusion

1. That what we refer to as mental events may be the same as quantum nonlocal events and spring from the same source. Two former mysteries can now be combined into one mystery, which can more readily be subjected to experiment in the biophysics realm.

2. It appears that macroscopic entanglement can be constantly regenerated or maintained by neuronal systems, while defying the usual decoherence faced by nonliving particle systems, permitting repeated measurements on these same systems. This allows one to introduce the concept of Alice and Bob being entangled themselves vs their previously just being limited to using entangled photon states. This also allows Alice and Bob, even if they are not entangled themselves, to use entangled human neuronal basins as their nonlocal proxies.

3. That we may now be able to test Hameroff's theory (1994) that microtubule based quantum coherence should be sensitive to general anesthesia using either entangled human subjects or human neuronal basins and, that by altering the neuronal microtubules in genetic or other fashion in the human neuronal basins, to possibly determine exactly where and how the transition from mental-quantum to brain-classical and vice-versa takes place.

4. That there appears to be two different types of wave-function collapse or decoherence processes, one for living and one for nonliving systems.

5. A possible experimental resolution of the Schrödinger's cat and Wigner's friend paradoxes is proposed which provides for a new nonlocal technique, based either upon entangled human subjects or human neuronal basins. This means that one never has to open up the box to see what has happened inside and can know instant by instant the status inside any closed box of any situations above and beyond the Schrődinger's cat and Wigner's friend paradoxes.

6. An empirical technique to determine if nonlocality is controllable or uncontrollable and whether superluminal communication is therefore possible in a 2-way fashion.

7. The possibility of simultaneously closing the detector efficiency and locality loopholes in a test of Bell's inequality by utilizing human neuronal basins as detectors.

**Acknowledgement**

I would like to thank the editor, the reviewers and Koichiro Matsuno for their valuable comments and suggestions. Also, Gary Shoemaker for answering my many difficult interdisciplinary questions and, once again, to Thesa von Hohenastenberg-Wigandt for first getting me started on this endeavor and for entanglement enlightenment

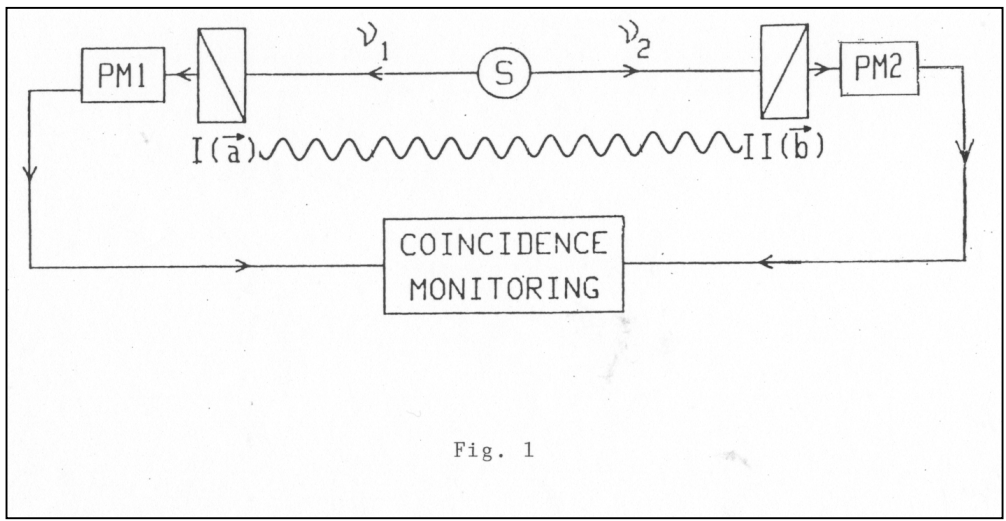

Fig. 1 Aspect EPR nonlocality experiment with photons (Reprinted with permission from the American Physical Society. Alain Aspect, Jean Dalibard, Gerard Roger. Phys. Rev. Lett. 49 (25), 1804 (1982). Copyright 1982 by the American Physical Society.

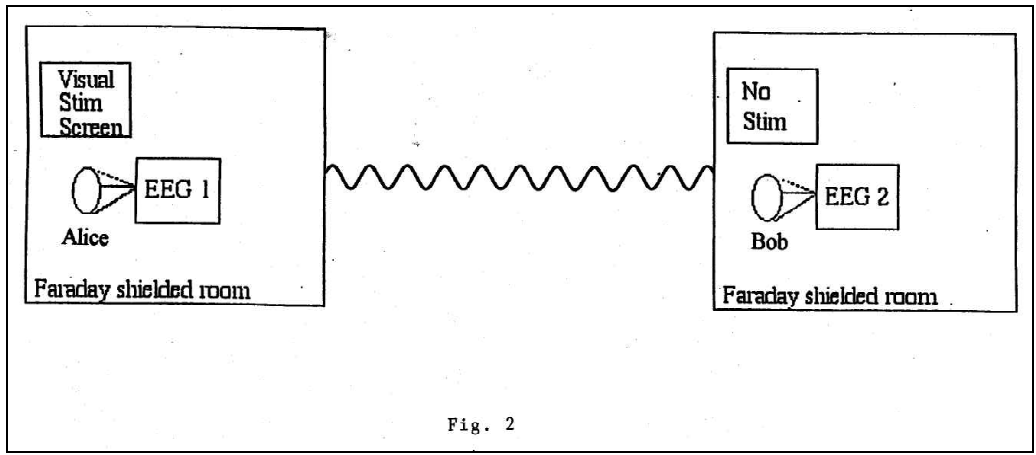

Fig. 2  Grinberg-Zylberbaum experiment with human subjects Alice and Bob inside Faraday cages

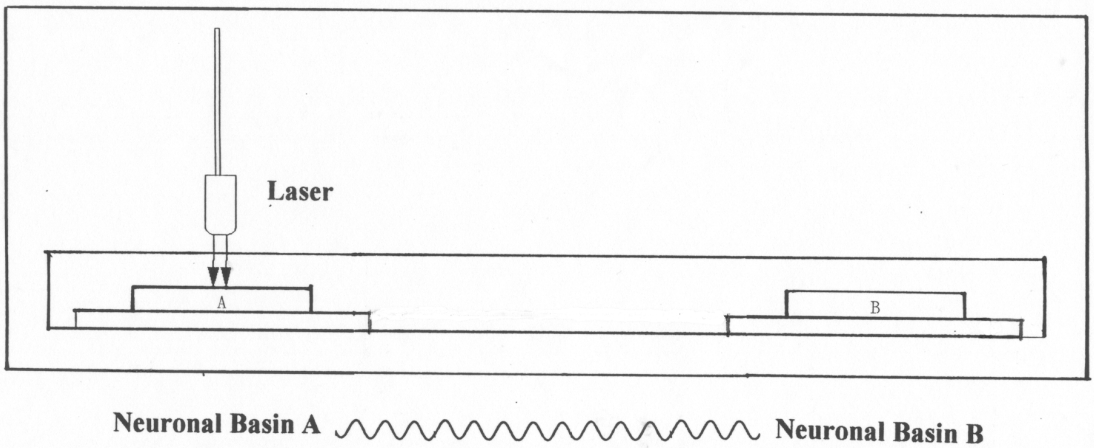

Fig. 3   Univ. of Milan human neuronal basins adhering to printed circuit boards and inside Faraday cages. The hypothetical wavy line represents nonlocal information transfer. (Reprinted with permission from the Dept. of Information Technologies, University of Milan.)

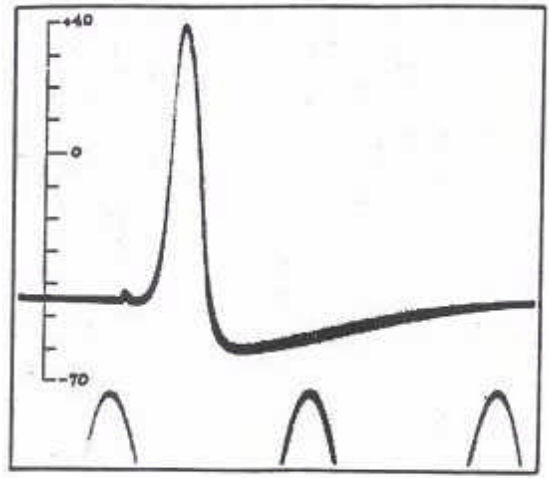

Fig. 4  Action potential with the vertical scale in millivolts and a time marker of 500 Hz. (Hodgkin, Huxley, 1939).

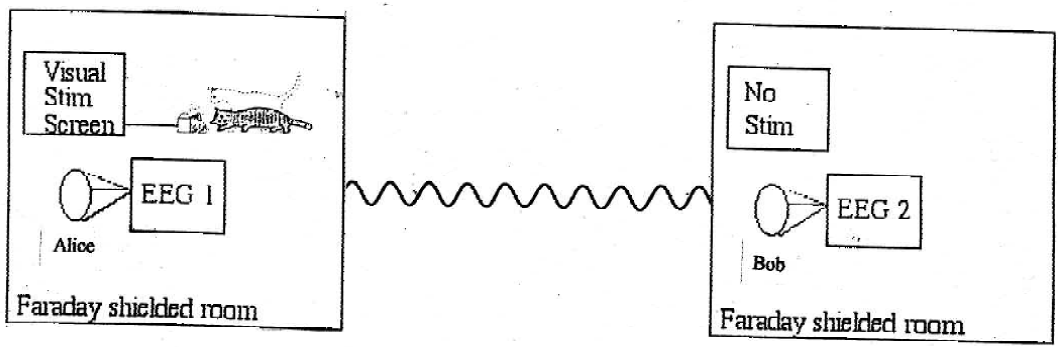

Fig. 5 Schrödinger's cat and Wigner's friend experiment

**References**

Aspect, A., Dalibard, J., Roger, G., 1982. Experimental test of Bell's inequalities using time-varying analyzers. Phys. Rev. Lett. 49, 1804-1807.

Atmanspacher, H. 2003. Mind and matter as asymptotically disjoint, inequivalent representations with broken time-reversal symmetry. BioSystems 68, 19-30.

Beck, F., Eccles, J.C. 1992. Quantum aspects of brain activity and the role of consciousness. Proc. Natl. Acad. Sci. USA. 89, 11357-11361.

Bell, J.S., 1964. On the Einstein-Podolsky-Rosen paradox. Physics 1, 195-200.

Bennett, C.H., Wiesner, S.J., 1992. Communication via one- and two-particle operators on Einstein-Podolsky-Rosen states. Phys. Rev. Lett. 69, 2881-2884.

Bennett, C.H., Brassard, G., Crepeau, C., Jozsa, R., Peres, A., Wooters, W.K., 1993. Teleporting an unknown quantum state via dual classical and Einstein-Podolsky-Rosen channels. Phys. Rev. Lett. 70, 1895-1899.

Bohr, N., 1958. Light and Life. In: Atomic Physics and Human Knowledge. Wiley, New York. Pp. 3-12.

Caldeira, A.O., Leggett, A.J., 1981. Influence of dissipation on quantum tunneling in macroscopic systems. Phys. Rev. Lett. 46, 211-214.

Chalmers, D.J., 1996. The conscious mind: In search of a fundamental theory. Oxford University Press, New York.

Conrad, M., Home, D., Josephson, B.D., Beyond quantum theory: a realist psycho-biological interpretation of physical reality. In: Tarozzi, G., van der Merwe, A., Selleri, F. (Eds.), Microphysical Reality and Quantum Formalism, vol. I. Kluwer Academic pub., Dodrecht., pp. 285-293.

Dietrich, A., Been, W., 2003. An extra dimensional approach of entanglement. quant-ph/0307117.

Eberhard, P.H., 1978. Bell's theorem and the different concepts of nonlocality. Il Nuovo Cimento 46B, 392-419.

Eccles, J.C., 1994. How the Self controls its Brain. Springer-Verlag, Berlin.

Einstein, A., Podolsky, B., Rosen, N., 1935. Can quantum mechanical description of physical reality be considered complete? Phys. Rev. 47, 777-780.

Ekert, A.K., 1991. Quantum cryptography based on Bell's theorem. Phys. Rev. Lett. 67, 661-663.

Freedman, S.J., Clauser, J.F., 1972. Experimental test of local hidden-variable theories. Phys. Rev. Lett. 28, 938-941.

Friedman, J.R., Patel, V., Chen, W., Tolpygo, S.K., Lukens, J.E., 2000. Quantum superpositions of distinct macroscopic states. Nature 406, 43-45.

Ghirardi, G., 1999. Quantum superpositions and definite perceptions: envisaging new feasible experimental tests. Phys. Lett. A 262, 1-14.

Goswami, A., 1993. The self-aware universe: how consciousness creates the material world. Penguin Putnam, New York. p.96.

Grinberg-Zylberbaum, G., Delaflor, M., Attie, L., Goswami, A., 1994. The Einstein-Podolsky-Rosen-Paradox in the brain: the transferred potential. Physics Essays 7, 422-428.

Hagan, S., Hameroff, S.R., Tuszynski, J., 2002. Quantum computation in brain microtubules; decoherence and biological feasibility. Phys. Rev. E 65, 061901.

Hameroff, S., 1994. Quantum coherence in microtubules: A neural basis for emergent consciousness? J. Consciousness Studies 1, 91-118.

Hameroff, S., 1998. Anesthesia, consciousness and hydrophobic pockets – a unitary quantum hypothesis of anesthetic action. Toxicology Lett. 100-101, 31-39.

Hodgkin, A.L., Huxley, A.F., 1939. Action potentials recorded from inside a nerve fibre. Nature 144, 710-711.

Hatori, K., Honda, H., Matsuno, K., 2001. Magnetic dipoles and quantum coherence in muscle contraction. quant-ph/0104042.

He, G-P, Zhu, S-L, Wang, Z.D., Li, H-Z., 2003. Testing Bell's inequality and measuring the entanglement using superconducting nanocircuits. quant-ph/0304156.

Josephson, B.D., Pallikari-Viras, F., 1991. Biological utilization of quantum nonlocality. Found. Phys. 21, 197-207.

Josephson, B.D., Rubik, B.A., 1992. The challenge of consciousness research. Frontier Perspec. 3 (1), 15-19.

Josephson, B.D., 2002. 'Beyond quantum theory: a realist psycho-biological interpretation of reality'revisited. BioSystems 64, 43-45.

Kwiat, P.G., Mattle, K., Weinfurter, H., Zeilinger, A., 1995. New high-intensity source of polarization-entangled photon pairs. Phys. Rev. Lett. 75, 4337-4341.

Leggett, A.J., 1987. Dynamics of the dissipative 2-state system. Rev. Mod. Phys. 59, 1-85.

Libet, B., 1994. A testable field theory of mind-brain interaction. J. Conscious. Stud. 1, 119-126.

Matsuno, K., 1993. Being free from ceteris paribus: a vehicle for founding physics on biology rather than the other way around. Appl. Math. Comput. 56, 261-279.

Matsuno, K., 1995. Quantum and biological computation. BioSystems. 35, 209-212.

Matsuno, K. 1999. Cell motility as an entangled quantum coherence. BioSystems, 51, 15-19.

Matsuno, K., Paton, R.C., 2000. Is there a biology of quantum information? BioSystems 55, 39-46.

Matsuno, K., 2001. The internalist enterprise on constructing cell motility in a bottom-up manner. BioSystems 61, 114-124.

Matsuno, K., 2003. Quantum mechanics in first, second and third person descriptions. BioSystems 68, 107-118.

Matsuno, K., 2004. Personal communication.

Mermin, N.D., 1998. What do these correlations know about reality? Nonlocality and the absurd. quant-ph/9807055.

Mitchell, M., Goswami, A., 1992. Quantum mechanics for observer systems. Physics Essays 5, 526-530.

Pearle, P., 1970. Hidden-variable example based upon data rejection. Phys. Rev. D. 2, 1418-1325.

Penfield, W., 1976. The mystery of the mind. Princeton University Press, Princeton.

Penrose, R., 1994. Shadows of the Mind. Oxford University Press, Oxford.

Pizzi, R., Fantasia, A., Gelain, F., Rosetti, D., Vescovi, A., 2003. Looking for quantum processes in networks of human neurons on printed circuit board. Quantum Mind 2, March 15-19 Tucson, Arizona. (http://www.consciousness.arizona.edu/quantum-mind2/abstracts.html) (http://www.dti.unimi.it/~pizzi).

Pizzi, R., Fantasia, A., Gelain, F. Rossetti, D., Vescovi, A., 2004a. Non-local correlation between human neural networks on printed circuit board. Toward a Science of Consciousness conference, Tucson, Arizona. (http://consciousness.arizona.edu/tucson2004) Abstract No. 104.

Pizzi, R., Fantasia, A., Gelain, F., Rossetti, D., Vescovi, A., 2004b. Nonlocal correlations between separated neural networks. Quantum Information and Computation II. ed. E. Donkor, A.R. Pirick, H.E. Brandt. Proceedings of SPIE 5436, 107-117.

Polchinski, J. 1991. Weinberg's nonlinear quantum mechanics and the Einstein-Podolsky-Rosen paradox. Phys. Rev. Lett. 66, 397-400.

Richards, T.L., Johnson, L.C., Kozak, L., King, H., Standish, L., 2002. EEG alpha wave evidence of neural energy transfer between human subjects at a distance. Tucson Toward a Science of Consciousness conference. (http://www.consciousness.arizona.edu/Tucson2002). Abstract No. 352.

Rowe, M.A., Kielpinski, D., Meyer, V., Sackett, C.A., Itano, W.M., Monroe, C., Wineland, D.J., 2001. Experimental violation of a Bell's inequality with efficient detection. Nature (London) 409, 791-793.

Schrödinger, E., 1935. Die gegenwärtige situation in der quantenmechanik. Naturwissenschaften 23, 807-812, 823-828, 844-849; 1980. The present situation in quantum mechanics. Proced. Am. Philosophical Soc. 124, 323-338. (John D. Trimmer, trans.) Reprinted in: Wheeler and Zurek, 1983. Quantum Theory and Measurement. Princeton Univ. Press, Princeton.

Shimony, A., 1963. Role of the observer in quantum theory. Am. J. Phys. 31, 755-773.

Shimony, A., 1984. Controllable and uncontrollable nonlocality. Proceedings of the International Symposium on Foundations of Quantum Theory. Phys. Soc. Tokyo, 130-139.

Standish, L., 2001. Neurophysiological measurement of nonlocal connectivity. In: Science and Spirituality of Healing conference. Kona, Hawaii. (available at: Samueli Institute, http://www.siib.org).

Standish, L.J., Johnson, L.C., Kozak, L., Richards, T., 2003. Evidence of correlated functional magnetic resonance imaging signals between distant human brains. Alter. Therapies 9 (1), 121-125.

Standish, L.J., Kozak, L., Johnson, L.C., Richards, T., 2004. Electroencephalographic evidence of correlated event-related signals between the brains of spatially and sensory isolated human subjects. J. Alter. Compl. Med. 10, 307-314.

Stapp, H.P., 1988. Are faster-than-light influences necessary? In: Quantum mechanics versus local realism-The Einstein-Podolsky-Rosen paradox. Ed. F. Selleri. Plenum Press, New York. pp. 63-85.

Stapp, H.P., 1994. Theoretical model of a purported empirical violation of the predictions of quantum theory. Phys. Rev. A. 50 (1), 18-22.

Stapp, H.P., 1997. Nonlocal character of quantum theory. Am. J. Phys. 65 (4), 301-304.

Tegmark, M., 2000. Importance of quantum decoherence in brain processes. Phys. Rev. E 61, 4194-4206.

Thaheld, F., 1998. Comments regarding Schrödinger's cat and Wigner's friend. Quantum-Mind Archives, Nov. 9. http://listserv.arizona.edu/archives/quantum-mind.html.

Thaheld, F., 1999a. The search for biological quantum nonlocality. Quantum-Mind Archives, July 16. http://listserv.arizona.edu/archives/quantum-mind.html.
43

Thaheld, F., 1999b. A testable nonlocal theory of mind-brain interaction. Quantum-Mind Archives, Aug. 10. http://listserv.arizona.edu/archives/quantum-mind.html.

Thaheld, F.H., 2000a. Proposed experiment to determine if there are EPR nonlocal correlations between two neuron transistors. Apeiron 7 (3-4), 202-205.

 (http://redshift.vif.com).

Thaheld, F., 2000b. Willful intent and biological nonlocality: a proposed paradigm with implications for healing intentionality. Proceedings: The Science and Spirituality of Healing Conference. available at: Samueli Institute, http://www.siib.org.

Thaheld, F.H., 2001a. A feasible experiment concerning the Schrödinger's cat and Wigner's friend paradoxes. Physics Essays. 14 (2), 164-170. www.physicsessays.com.

Thaheld, F.H., 2001b. A preliminary indication of controllable biological quantum nonlocality? Apeiron 8 (1). (http://redshift.vif.com).

Thaheld, F.H., 2003a. Can we determine if the linear nature of quantum mechanics is violated by the perceptual process? BioSystems 71, 305-309.

Thaheld, F.H., 2003b. Biological nonlocality and the mind-brain interaction problem: comments on a new empirical approach. BioSystems 70, 35-41.

Thaheld, F.H., 2004a. The potential role of biologically entangled neuronal basins in the resolution of various problems in the field of consciousness studies. Toward a Science of Consciousness Conference. Tucson, Arizona. (http://consciousness.arizona.edu/tucson2004).

Thaheld, F.H., 2004b. A method to explore the possibility of nonlocal correlations between brain electrical activities of two spatially separated animal subjects. BioSystems 73, 205-216.

Tittel, W., Brendel, J., Zbinden, H., Gisin, N., 1998. Violation of Bell inequalities by photons more than 10 km apart. Phys. Rev. Lett. 81, 3563-3566.

Uzan, J., Luminet, J., Lehoucq, R., Peter, P., 2000. Twin paradox and space topology. physics/0006039.

Vaidman, L., 2001. Bell's inequality: more tests are needed. quant-ph/0102139.

von Neumann, J., 1955. Mathematical Foundations of Quantum Mechanics. Princeton Univ. Press, Princeton.

Wackermann, J., Seiter, C., Keibel, H., Walach, H., 2003. Correlations between brain electrical activities of two spatially separated human subjects. Neurosci. Lett. 336, 60-64.

Walter, W.G., 1950. Electroencephalography-A Symposium on its Various Aspects. McMillan, New York. p.203.

Weihs, G., Jennewein, T., Simon, C., Weinfurter, H., Zeilinger, A., 1998. Violation of Bell's inequality under strict Einstein locality conditions. Phys. Rev. Lett. 81, 5039-5043.

Weiss, U., Grabert, H., Linkwitz, S., 1987. Influence of friction and temperature on coherent quantum tunneling. J. Low Temp. Phys. 68, 213-244.

Whitaker, A., 1996. Einstein, Bohr and the Quantum Dilemma. University Press, Cambridge.

Wigner, E.P., 1961. in: The Scientist Speculates-An Anthology of Partly Baked Ideas. ed. L.J. Good. Heinemann, London.

Wigner, E.P., 1963. The problem of measurement. Am. J. Phys. 31, 6-15.